\documentclass[twoside]{article}
\usepackage{amsbsy}
\usepackage{amsfonts}
\usepackage[lmargin=1in,rmargin=1in,tmargin=1in,bmargin=1in]{geometry}
\usepackage[numbers]{natbib}

\usepackage{graphicx}
\usepackage{abstract}
\usepackage{caption}
\usepackage{ragged2e}
\usepackage{sectsty}
\subsectionfont{\fontsize{10}{10}\normalfont\itshape}
\sectionfont{\fontsize{10}{10}\normalfont\bfseries}

\usepackage{fancyhdr}


\date{}

\begin{document}

\title{\textbf{Flexible model selection\\for mechanistic network models}}

\author{ \textsc{Sixing Chen}\\ \textit{Department of Biostatistics, T.H. Chan School of Public Health, Harvard University}\\
\textit{655 Huntington Avenue, Building 2, 4th Floor, Boston, MA, USA 02115}
\and \textsc{Antonietta Mira}\\ \textit{Data Science Lab, Institute of Computational Science, Universit\`a della Svizzera italiana}\\
\textit{Via Buffi 6, 6900 Lugano, Switzerland} \\
\textit{and Dipartimento di Scienza e Alta Tecnologia, Universit\`a degli Studi dell'Insubria}\\
\textit{Via Valleggio, 11 - 22100 Como, Italy}\\
\textsc{and}\\
\and \textsc{Jukka-Pekka Onnela$\dag$}\\ \textit{Department of Biostatistics, T.H. Chan School of Public Health, Harvard University}\\
\textit{655 Huntington Avenue, Building 2, 4th Floor, Boston, MA, USA 02115}\\
$\dag$ Corresponding author. Email: onnela@hsph.harvard.edu}

\maketitle
\thispagestyle{empty}

\begin{abstract}
{Network models are applied across many domains
where data can be represented as a network. Two
prominent paradigms for modeling networks are statistical models (probabilistic
models for the observed network) and mechanistic models (models
for network growth and/or evolution). Mechanistic models are
better suited for incorporating domain knowledge, to study effects of interventions (such as changes to specific mechanisms)
and to forward simulate, but they typically have intractable likelihoods.
As such, and in a stark contrast to statistical models, there is a
relative dearth of research on model selection for such models despite the otherwise
large body of extant work. In this paper, we propose a simulator-based
procedure for mechanistic network model selection that borrows aspects
from Approximate Bayesian Computation (ABC)
along with a means to quantify the uncertainty in the selected model.
To select the most suitable network model, we consider and assess the performance
of several learning algorithms, most notably the so-called Super Learner,
which makes our framework less sensitive to the choice of a particular learning algorithm.
Our approach takes advantage of the ease to forward simulate from
mechanistic network models to circumvent their intractable likelihoods.
The overall process is flexible and widely applicable.
Our simulation results demonstrate the approach's ability to accurately
discriminate between competing mechanistic models. Finally, we
showcase our approach with a protein-protein interaction network model from the literature for yeast
(\textit{Saccharomyces cerevisiae}).}

\end{abstract}

\section{Introduction}

Many systems of scientific and societal interest can be represented
as networks, and network models are used, among many other application areas,
to study social networks, communication patterns, scientific citations,
and protein-protein interactions \citep{wasserman1994social,pastor2007evolution,mej2010networks,lusher2012exponential,NetworkExpRaval2013introduction}.
There are (at least) two prominent paradigms to the modeling of networks,
which we refer to as the statistical approach and the mechanistic approach.
In the statistical approach, one describes a model that specifies
the likelihood of observing a given network, i.e., these are probabilistic
models of data that take the shape of a network \citep{robins2007introduction,LatSpaceHoff2002latent,goyal2014sampling}.
In the mechanistic approach, one specifies a collection of domain-specific
microscopic mechanistic rules, informed by scientific understanding
of the problem, that are used to grow or evolve the network over time
\citep{barabasi1999emergence,watts1998collective,sole2002model,vazquez2003modeling,klemm2002highly,kumpula2007emergence}.
Both modeling approaches provide distinct advantages.

In mechanistic models, a particular generative mechanism may seem
like a strong assumption in a context where one does not directly observe
the formation of the network, and where it is difficult to study microscopic
interactions in isolation. For example,
it might be difficult to learn about the mechanistic rules that govern
the formation or dissolution of ties in in-person interactions, whereas
doing so in an online social network or a phone communication
network \citep{viswanath2009evolution,onnela2007structure,saramaki2015seconds},
where every interaction can be recorded, is feasible. In
some biological networks the pairwise interactions between the elements
are well understood both theoretically and experimentally, they can
be studied in isolation, and these interactions are reproducible.
For example, gene duplication is one of the main drivers of the evolution
of genomes, it is well understood, and therefore perhaps not surprisingly,
network models based on gene duplication were one of the first large-scale
models used in systems biology \citep{NetworkExpRaval2013introduction}.

In comparison, common statistical models
have limitations in the structures they are able to accommodate \citep{goyalManu},
and fitting and sampling from some of these models can be difficult.
For example, the popular class of exponential random graph models
(ERGMs) may not always be consistent under sampling \citep{shalizi2013consistency}.
Mechanistic models do not suffer from these limitations as much, since
generation of network structures from a handful of mechanisms is usually
computationally inexpensive, so it is relatively simple to sample or forward
simulate (to simulate observations given parameter values) from a
particular model. Another advantage of mechanistic models is the ease with which one
can incorporate domain knowledge in the model. Since the modeler is
in control of the mechanisms to include, one is able to encode relevant
domain knowledge of known or hypothesized interactions between actors
in the system as mechanistic rules. The duplication-divergence models
for protein-protein interaction networks are good examples of this
\citep{NetworkExpRaval2013introduction}. Additionally, it is easy to study the
effects of interventions on the behaviors of the actors by directly modifying
the model's mechanistic rules to reflect the intervention \citep{wang2014sample,goyal2014sampling}.

While there is an extensive literature on mechanistic models in
network science, there is a comparative dearth of work on model selection in mechanistic
models \citep{middendorf2005inferring,ratmann2009model,thorne2012graph,onnelaManu}. The
aim of this paper is to provide a general framework for model selection for
mechanistic network models. For instance, given a full model which
has an array of different generative mechanisms,
we are interested in selecting between different submodels each possessing
only a subset of the mechanisms of the full model. Traditional likelihood-based
model selection, frequentist or Bayesian, is not applicable to mechanistic
models because in most cases their likelihood functions are intractable.
The main reason for this is that in mechanistic models one must consider all the
possible paths to generate any one particular network realization,
which leads to a combinatorial explosion save for the most trivial
settings. As such, one must resort to likelihood-free approaches.

A recent likelihood-free approach to both inference and model selection
for problems with intractable likelihoods is Approximate Bayesian
Computation (ABC) \citep{Marin2012,sunnaaker2013approximate,lintusaari2017fundamentals,sisson2018handbook}.
This Bayesian approach aims at calibrating the
model parameters to obtain an approximate posterior distribution for
the parameters of interest for a given observed data set. Following Bayes theorem, the posterior
is obtained by combining information from the prior distribution on
parameters and the observed data set as incorporated by the likelihood. ABC
inference starts by generating samples of possible parameter values
from the prior. For each sample, one forward simulates
a data set from the model, where the nature
of the model, statistical or mechanistic, is not relevant. Then, given
a distance measure between the observed data (the target of inference and/or model selection) and the model generated data,
one accepts only those data sets from the model that are within a certain distance from the observed data.
The parameters sampled from the prior corresponding to these accepted data
sets form an approximation to the posterior distribution. Unless
the data are discrete and of low dimension, it is usually necessary
to base the distance measure on summary statistics of the data \citep{beaumont2010approximate}.
Model selection with ABC is similar but includes an additional layer
of hierarchy and a prior for the candidate model indices \citep{grelaud2009abcMC}.
More generally, there is a scarcity of likelihood-free model selection methods
outside the realm of ABC. \citet{toni2009approximate} extend the
sequential Monte Carlo ABC of \citet{sisson2007sequential} to ABC
model choice, while \citet{lee2015model} do so with the approach
of \citet{fearnhead2012constructing}. Other recent innovations include
an adaptive approach \citep{stoehr2015adaptive} and a random forest-based
approach \citep{RF_ABCpudlo2015reliable}.

The main difficulty in practice for ABC arises when selecting the summary
statistics as well as the threshold on the distance. In general, any quantity
that can be  calculated from the observed data is called a statistic;
if the likelihood depends on the data only through some low-dimensional statistic,
so that we need only this statistic to calculate the likelihood for any value of the parameter,
the statistic is said to be sufficient. In the context of ABC, if the selected summary
statistics are sufficient for the parameters of the model and the
distance threshold is zero, i.e., only parameter values that have given rise to
data sets with values of sufficient summary statistics exactly matching those
of the observed data set are retained, then these accepted parameter
samples will be from the true posterior \citep{barber2015rate,lintusaari2017fundamentals}.
However, should one fall short on either of these, then the
accepted parameter samples will only be from an approximation to the
true posterior. When available, relevant domain knowledge can be applied
to guide the selection of ``essentially sufficient'' summary statistics,
which capture features attributable to the inferential objects
of interest. However, since likelihood-free approaches like ABC are
only needed with analytically or computationally intractable likelihoods,
it will typically be difficult to find the sufficient statistics in
the absence of domain knowledge, though there is previous work on
how to select good summary statistics for an ABC procedure in such
settings \citep{drovandi2011approximate,ABCstatCHOOSEprangle2014semi}.
As for the threshold on distance, the smaller the distance
threshold, the lower the acceptance rate, and hence the greater the
computational burden to generate a reasonable number of accepted samples.
In fact, outside of using discrete summary statistics, it may be
impractical to use a distance threshold of zero. As a result, the
performance of ABC inference can suffer due to the inaccuracy of the
resulting posterior \citep{sisson2007sequential}.

ABC model choice
suffers from these same issues for to the model index,
which becomes an additional model parameter on which inference is
required. Even if one were to select statistics that are marginally
sufficient for the candidate models, they may not be jointly sufficient for
the overall model (the overall hierarchical model that is indexed by the candidate
model index and the corresponding parameters of each candidate model)
and thus may not be able to discriminate among the various
models under comparison, save for some special cases \citep{grelaud2009abcMC,BFprobRobert2011lack}.
Lastly, even if one does manage to select all the sufficient statistics
and conduct model selection with ABC, there may be a large discrepancy between the
resulting ABC Bayes factor and the true Bayes factor \citep{BFprobRobert2011lack}.

Instead of dealing with issues stemming from the inaccuracy of
the ABC posterior and Bayes factor, we propose a procedure for model
selection that borrows from ABC the aspect of data generation from candidate models.
Just as in ABC model choice, the data forward simulated from
each candidate model will be the basis for model selection as it becomes
the training data, but rather than using a full Bayesian
approach, we propose to conduct model selection with a flexible
learning algorithm. We assess the performance of different learning algorithms,
including the Super Learner
(SL) \citep{SLbookPolley2011super,SLvan2007super}. Originally proposed
for prediction in a regression setting, SL is an ensemble learning algorithm
that makes a prediction by combining the predictions from a library
of candidate algorithms. Given a particular loss function, SL aims
to minimize the expected loss, called the risk. The \textit{discrete SL} simply
picks the candidate algorithm that has the lowest cross-validated
risk over the training data, whereas the \textit{full SL} creates the convex
combination of the candidate algorithm-specific estimates that has the lowest
cross-validated risk. Given a bounded loss function, both the discrete
and the full SL have the so-called oracle property, meaning that asymptotically
(in the number of training data sets) they perform at least as well
as the optimal candidate algorithm available in the library and as
the optimal convex combination of the candidate algorithms, respectively
\citep{ORACLEvan2003unified,ORACLEdudoit2005asymptotics}. Due to the intractable nature of the likelihood and the ease of forward
simulation, model selection with mechanistic network models lends
itself well to the flexibility of the SL.

Our proposed approach shares similarities with that of \citet{RF_ABCpudlo2015reliable},
which is a random forest-based ABC approach for model selection that
is fairly robust to the choice of summary statistics. Their approach
measures performance with the prior error rate, which is the probability
to select the wrong model averaged over the prior. In contrast, in our approach with the SL,
the choice of performance measure is flexible and can be encoded directly
into the loss function. For example, the prior error rate implicitly
weighs misclassification differently for each model due to the sensitivity
to the choice of the prior. Should one desire a measure that does
not discriminate between misclassification of different models, one
can use a measure like the Area Under the receiver operating characteristic
Curve ($AUC$) \citep{AUCbradley1997use,AUCling2003auc}. Furthermore,
SL can make use of a host of candidate algorithms, including random
forest, to perform the classification. Random forest is a very flexible learning
algorithm, but it may not perform well in all settings. One can be
robust against this by having more candidate algorithms in the
SL library. To handle uncertainty
in the selected model, we adapt the regression-based approach used
by \citet{RF_ABCpudlo2015reliable}.

The rest of the
paper is organized as follows. In Section 2, we provide a brief overview
of SL as well as the procedure for model selection in the context
of mechanistic network models. We also introduce and motivate a simple
mechanistic network model as a proof of concept
in our subsequent simulations. In Section 3, we lay out the
details of the simulations as well as the results, and evaluate the
performance of our approach. In Section 4, we apply our approach to the
selection between two candidate models for a protein-protein
interaction data set. Finally, in Section 5, we conclude with
discussions and suggestions for future work.

\section{Methods and Materials}

\subsection{Overview of Super Learner}

The Super Learner (SL) is a fairly recent ensemble machine learning method
introduced by \citet{SLvan2007super}. Given a particular loss function $L$, the SL 
aims to minimize the expected loss $E\left[L\right]$, known as the
risk, with respect to the distribution of the training data with a
prediction algorithm composed of a library of candidate algorithms
$\left\{ Q_{l}\right\}$, such as random forest and support vector machine. The procedure begins by partitioning the
training data, with predictors $\boldsymbol{X}_{t}$, and outcome
$\boldsymbol{Y}_{t}$, into $V$ validation sets. In our setting, when applied to
model selection for mechanistic network models, summary statistics
computed on forward simulated network realizations play the role of
predictors and network model indices play the role of outcome
in SL. The predictors and outcome of the $v$th validation set are
referred to as $\boldsymbol{X}_{t}^{v}$ and $\boldsymbol{Y}_{t}^{v}$,
respectively, while those of the corresponding training set, i.e.,
the union of the remaining $V-1$ validation sets, are referred to
as $\boldsymbol{X}_{t}^{-v}$ and $\boldsymbol{Y}_{t}^{-v}$. Note
the distinction between training data ($\boldsymbol{X}_{t}$ and $\boldsymbol{Y}_{t}$)
and training set ($\boldsymbol{X}_{t}^{-v}$ and $\boldsymbol{Y}_{t}^{-v}$).
For the $v$th validation set, each candidate algorithm $Q_{l}$ of
the library is trained on $\left(\boldsymbol{X}_{t}^{-v},\boldsymbol{Y}_{t}^{-v}\right)$.
The resulting trained candidate algorithm $\hat{Q}_{l}^{v}$ is then evaluated
at $\boldsymbol{X}_{t}^{v}$, giving prediction $\hat{\boldsymbol{Y}}_{l}^{v}$.
After training each candidate algorithm on each training set and evaluating
it on the corresponding validation set, a new data set $\boldsymbol{Z}=\{ \{ \hat{\boldsymbol{Y}}_{l}^{v}\} ,\boldsymbol{Y}_{t}^{v}\} $
is formed with the cross-validated predicted outcomes $\{ \hat{\boldsymbol{Y}}_{l}^{v}\}$
generated using all candidate algorithms in $\left\{ Q_{l}\right\}$
and the true outcome $\boldsymbol{Y}_{t}^{v}$ from all validation
sets, where the former serve as the new predictor and the latter
as the new outcome. $\boldsymbol{Z}$ is used to estimate the
cross-validated risk. This cross-validation procedure is intended
to prevent overfitting, and this new data set will be the basis for
the final prediction algorithm.

\begin{figure}[!h]
\centering\includegraphics[width=\textwidth]{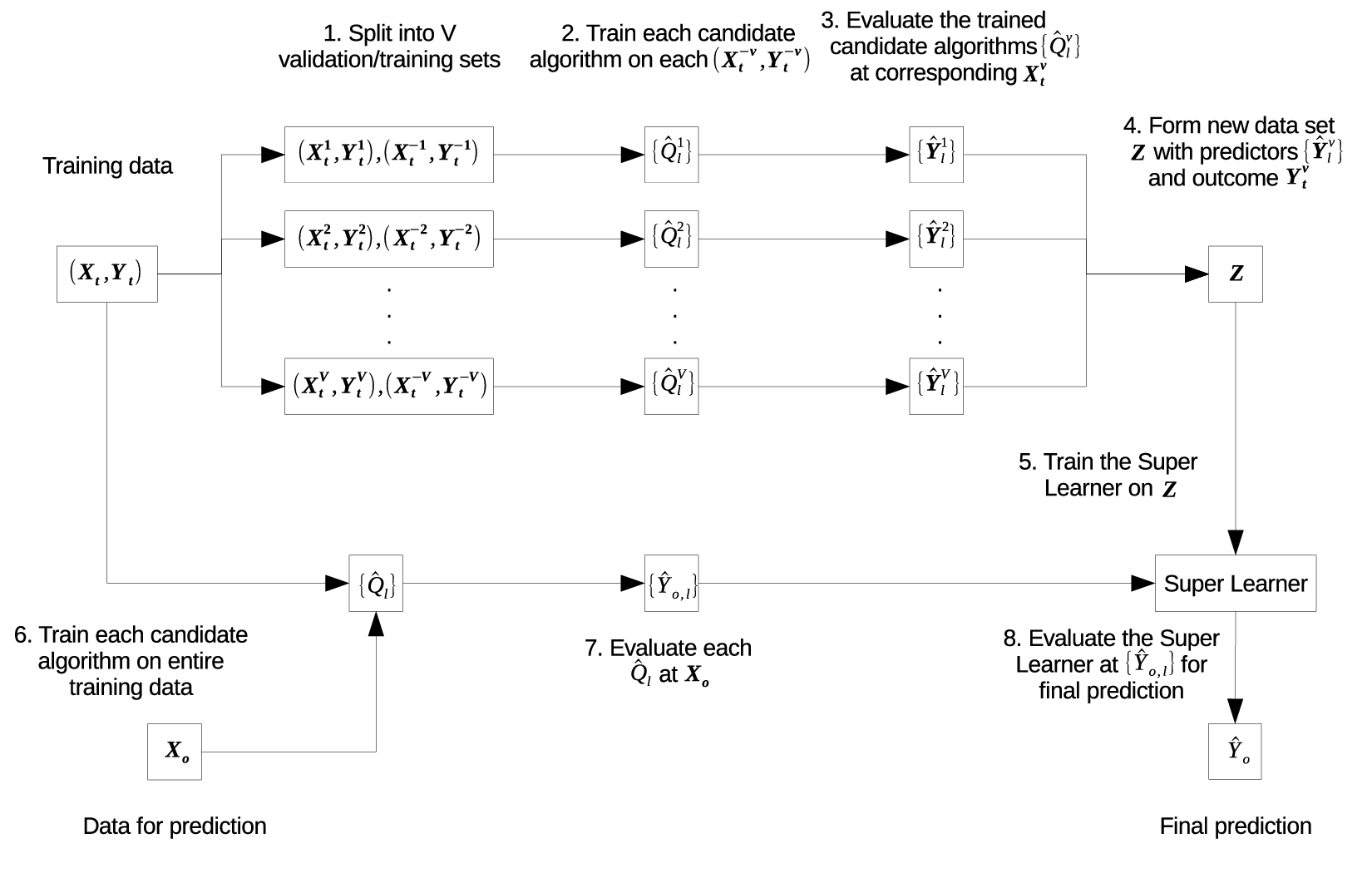}
\caption{Schematic of the SL framework. Note that $\boldsymbol{X}_{t}$ and $\boldsymbol{Y}_{t}$ are the model generated training data for network model selection, while $\boldsymbol{X}_{o}$ are the data from the observed network.}
\label{fig_1}
\end{figure}

To arrive at the final prediction via $\boldsymbol{Z}$, in the case
of the discrete SL, the candidate algorithm with the smallest estimated
cross-validated risk is chosen for final prediction. Assuming a regression
setting and a squared error loss function, the cross-validated risk
for the $l$th candidate algorithm can be estimated as
\[\hat{E}\left[L\left(Q_{l}\right)\right]=\frac{1}{V}\sum_{v}\frac{1}{n_{v}}\sum_{i}(Y_{t,i}^{v}-\hat{Y}_{l,i}^{v})^{2},\]
where the first summation is over the $V$ validation sets and the
second is over the $n_{v}$ observations in the $v$th validation
set. In the case of the full SL, the estimated risk is minimized
over all convex combinations of the candidate algorithms. In the same
regression setting with squared loss and a particular convex combination
$\boldsymbol{a}$, where $\sum_{l}a_{l}=1$ and $a_{l}\ge0$, the
cross-validated risk is estimated as
\[\ensuremath{\hat{E}\left[L\left(\boldsymbol{a}\right)\right]=\frac{1}{V}\sum_{v}\frac{1}{n_{v}}\sum_{i}(Y_{t,i}^{v}-\sum_{l}a_{l}\hat{Y}_{l,i}^{v})^{2}}.\]
Once the final prediction algorithm is determined, i.e., $Q_{l^{*}}$
that achieves the smallest risk in the discrete SL or $\boldsymbol{a}^{*}$
in the full SL, each candidate algorithm is refit on the entire training
data in order to predict an outcome for $\boldsymbol{X}_{o}$, the
observed predictors. Each resulting trained candidate algorithm $\hat{Q}_{l}$
is then evaluated at $\boldsymbol{X}_{o}$, giving prediction $\hat{Y}_{o,l}$.
The final prediction is $\hat{Y}_{o,l^{*}}$ in the discrete SL,
or $\sum_{l}a_{l}^{*}\hat{Y}_{o,l}$ in the full SL. In the classification
setting, such as our model selection framework, the $\{ \hat{\boldsymbol{Y}}_{l}^{v}\} $ are
the candidate algorithm-specific scores for each class and $\boldsymbol{Y}_{t}^{v}$
the true class, with $\hat{E}\left[L\left(Q_{l}\right)\right]$ and
$\hat{E}\left[L\left(\boldsymbol{a}\right)\right]$ defined according
to the chosen loss function. The final prediction $\hat{Y}_{o,l^{*}}$
or $\sum_{l}a_{l}^{*}\hat{Y}_{o,l}$ are then the scores for each
potential class. Figure \ref{fig_1} gives a visual representation of
the SL framework.

\subsection{Model Selection Framework}

Next, we introduce the proposed procedure for mechanistic network
model selection within the SL framework. In this setting, the SL will be based on
training data forward simulated from each candidate mechanistic network
model in order to predict the model index (outcome) from a set of
chosen network statistics (predictors). Before going ahead with the SL procedure,
one needs to determine the appropriate loss function.
Since the prediction is for the model index, this is a classification
problem, and therefore a loss function like squared loss is no longer appropriate.
Instead, we propose to use $L_{AUC}$ as the loss function.
$L_{AUC}$, also known as ``rank loss,'' is the loss function associated
with $AUC$, the area under the receiver operating characteristic
curve. The $AUC$ is an appropriate measure of the quality of the
classification since it does not depend on the distribution of the
model index in the data for performance evaluation. The corresponding
loss function is bounded, so the resulting SL will retain the oracle
property \citep{ORACLEvan2003unified,ORACLEdudoit2005asymptotics}.

\begin{figure}[!h]
\centering\includegraphics[width=\textwidth]{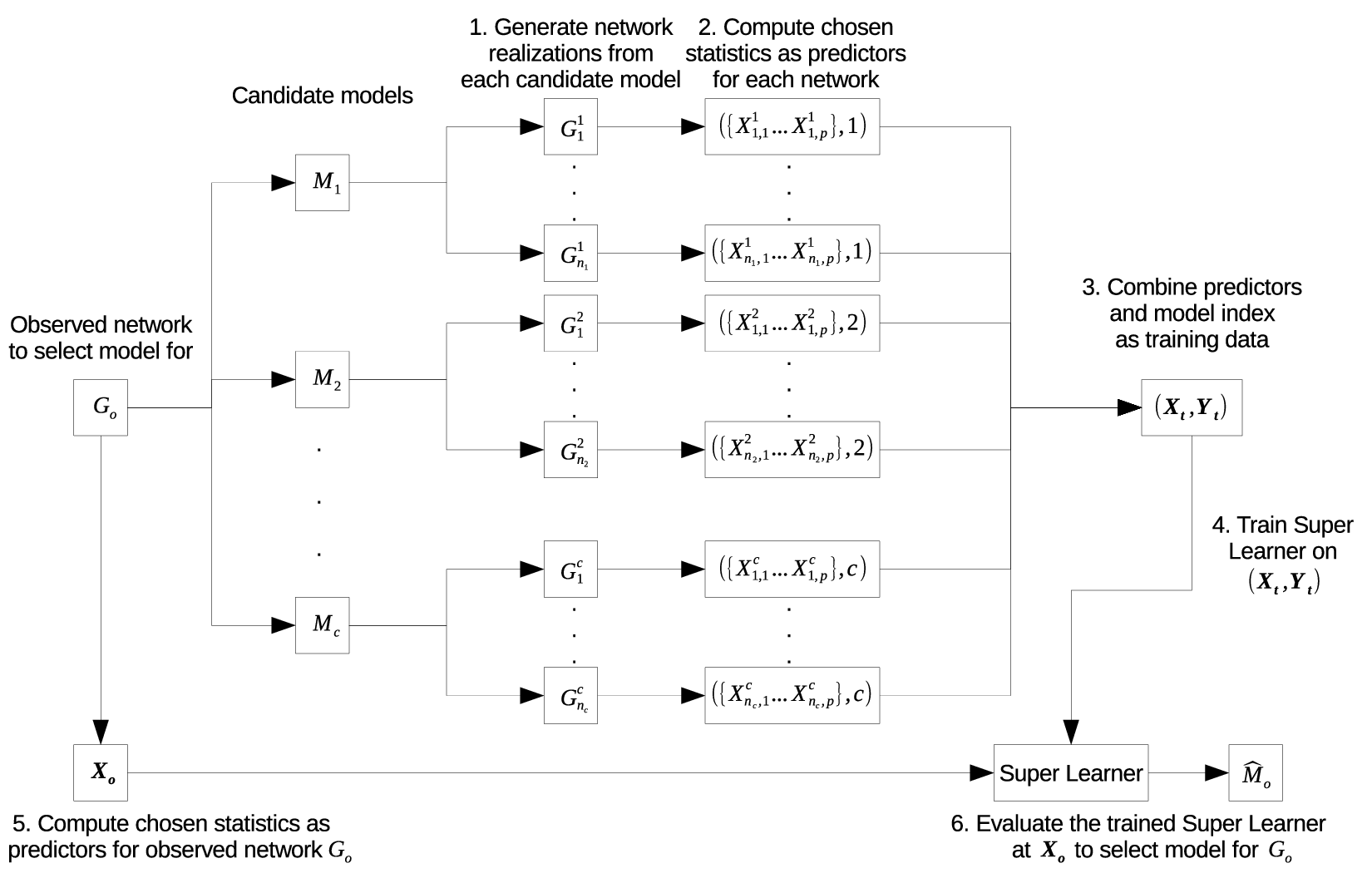}
\caption{Schematic of model selection procedure with SL. Note that steps 1-6 of Fig. \ref{fig_1} are contained in step 4 in Fig. \ref{fig_2}.
The arrow between $G_o$ and the candidate models is merely to indicate the influence of the empirical network on the candidate models, i.e., through model fitting.}
\label{fig_2}
\end{figure}

\textbf{Algorithm 1} lays out the procedure we propose for model selection
using SL, and Fig. \ref{fig_2} is the corresponding schematic. In addition
to selecting the candidate algorithms, one needs to select the all-important
network summary statistics to train the candidate algorithms on, in steps 4 and 6 in \textbf{Algorithm
1}. As previously stated in \citet{BFprobRobert2011lack}, even if
the sufficient statistics of all candidate models are selected, they may
not be jointly sufficient for the overall model. Since it will be difficult
or impossible to achieve, sufficiency cannot be a
criterion in choosing these statistics, but rather their ability to
characterize the differences between data generated by the candidate
models and thus their ability to discriminate among the models. Suppose
one is trying to select between a full model and one of its submodels
that has one of the mechanisms of the full model removed or ``turned off.'' In this
case, one needs to consider the characteristics of the network that
the missing mechanism affects. The statistics chosen as predictors in steps 4 and 6
should reflect the characteristics affected by the differences in the model,
either the different mechanisms themselves or different parameter values of the included mechanisms.
For instance, should the submodel be missing the triadic closure mechanism, which refers to a
mechanism that ``closes'' a triad of three nodes and two edges by adding the third edge,
one would expect statistics related to clustering to be affected. The ability to characterize these
differences will determine the performance of the overall procedure and,
thus, should guide the selection of the summary statistics. Though
the candidate models we consider here are nested, they do not need
to be in general to use this framework. Finally, in step 7 of \textbf{Algorithm
1}, evaluating the trained SL on the predictors for the observed network
gives a score to classify the observation. In binary classification,
i.e., having two candidate models, one model will be nominally ``negative''
and represented by 0 in the training of the SL, while the other will
be nominally ``positive'' and represented by 1. In this case, the
score will be bounded between 0 and 1 with scores above a user-defined
cutoff classified as 1, or the ``positive'' model, and vice versa.
For example, a cutoff of 0.5 can be used for this binary situation,
as we did in our simulations.  Note that due to the insensitivity of the $AUC$
to the model index prior in the training data, no prior should be
placed on the model index. In contrast, priors for parameters of candidate models
should be considered.

\begin{description}
\item[\textbf{Algorithm 1}] Steps for proposed mechanistic network model selection via SL:
\end{description}

\begin{enumerate}
\item Select relevant statistics that \textbf{highlight differences} between models
as predictors
\item Generate training data from all models of interest
\item Split the training data into cross-validation sets
\item Train/evaluate each candidate algorithms on each training/validation
pair based on selected predictors
\item Train SL on the results from each candidate algorithm
\item Train each candidate algorithm on the entire training data set
\item Classify/select model for observed network based on the results from
steps 5 and 6 of this algorithm using a preselected cutoff on the score.
\end{enumerate}

In order to perform uncertainty quantification, i.e., to quantify
the confidence in the model selected through \textbf{Algorithm 1},
we adapt a procedure for the same purpose from \citet{RF_ABCpudlo2015reliable}.
In \citet{RF_ABCpudlo2015reliable}, for each element in the training
data, one needs to produce an out-of-bag (OOB) classifier, i.e., a
random forest classifier based only on trees that do not involve (were not fitted using)
the given element. OOB shares similarities with cross-validation,
which SL does in steps 1-3 in Fig. \ref{fig_1}, but step 4 in Fig.
\ref{fig_1} involves the use of the whole data set. This means that the base SL is not completely cross-validated. 
To overcome this, we propose to first split the training
data into $B$ subsets $(\boldsymbol{X}_{t}^{b},\boldsymbol{Y}_{t}^{b})$, and the classifier for any element in $(\boldsymbol{X}_{t}^{b},\boldsymbol{Y}_{t}^{b})$
is defined as the SL trained on $(\boldsymbol{X}_{t}^{-b},\boldsymbol{Y}_{t}^{-b})$,
i.e., the training data without $(\boldsymbol{X}_{t}^{b},\boldsymbol{Y}_{t}^{b})$.
For each element of the training data, with $I$ denoting the indicator
function, we compute $W=I(\widehat{M}^{*}=Y)$, where
$\widehat{M}^{*}$ is the model selected by the corresponding 
classifier and $Y$ is the true model index. Now, one can build a
binary regression model (since $W$ is a binary indicator for correct model classification)
for $P(\widehat{M}^{*}=Y|\boldsymbol{X})$
by regressing $\boldsymbol{W}_{t}$, i.e., $W$ computed for all elements
of the training data, on the predictors $\boldsymbol{X}_{t}$. This
regression model can be simply logistic regression or a SL with the
correctly specified loss function. Lastly, one can take the fitted
value of this regression model at $\boldsymbol{X}_{o}$, the predictors
for the observed network, as the estimate for $P(\widehat{M}_{o}^{*}=Y_{o}|\boldsymbol{X}_{o})$,
where $Y_{o}$ is the unobserved true model index for the observed
network.

\section{Simulation Study}

\subsection{Variant of the Erd\H{o}s-R\`enyi Model}

As a proof of concept for our framework for mechanistic model selection,
we introduce a simple mechanistic model.
The basis for the model is the classic Erd\H{o}s-R\`enyi
(ER) model \citep{ERerdds1959random}. In the ER model, the number
of nodes $n$ is fixed, and there are two variants on how edges are
placed in the graph. In one variant, sometimes called the $G(n,p)$
model, each of the $C(n,2)$, $n$ choose 2, possible edges
are independent and included in the graph with probability $p$, so
the number of edges in the graph has a binomial distribution. In the
other variant, sometimes called the $G(n,m)$ model, the number of
edges in the graph $m$ is also fixed. In this case, the random graph
has a uniform distribution over all $C(C(n,2),m)$
possible graphs with $n$ nodes and $m$ edges. 

\begin{figure}[!h]
\centering\includegraphics[width=4in]{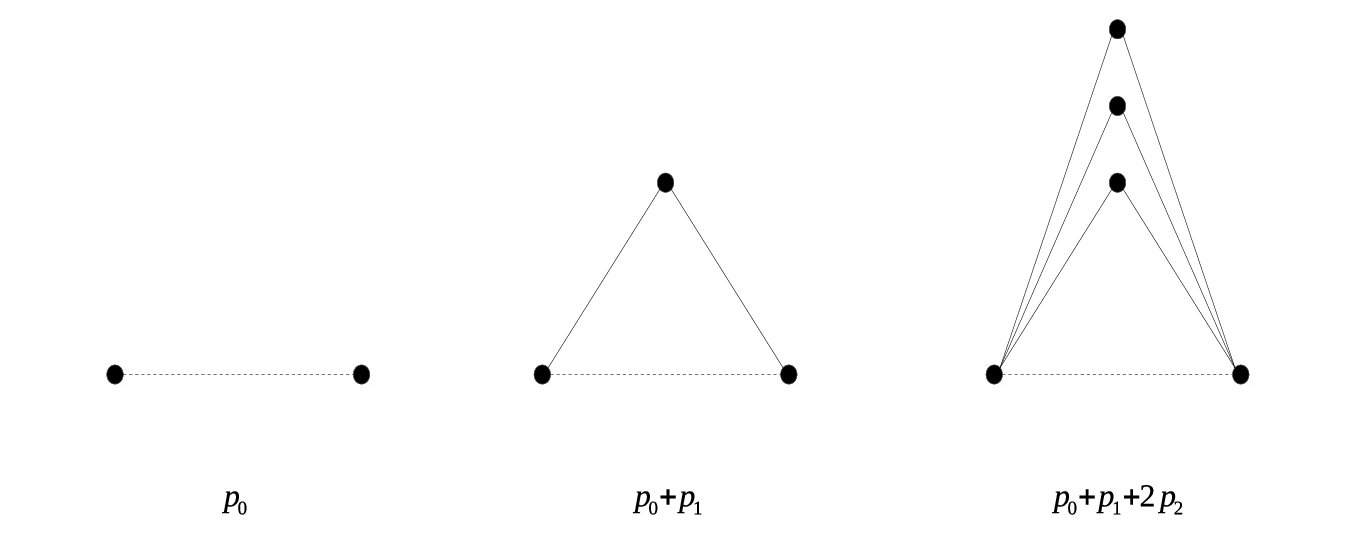}
\caption{The probability to add an edge between
two unconnected nodes in our variant of the ER model with two additional
mechanisms when there are no closeable triads (left), when there
is one closeable triad (middle), and when there is more than one
closeable triad (right). Here $p_{0}$ is the base probability for edge
placement, $p_{1}$ is the increase in the probability for closing
at least one triad, and $p_{2}$ is the increase for each additional
closeable triad after the first.}
\label{fig_3}
\end{figure}

Our model takes elements from both variants of the ER model. The model
generates random graphs with a fixed number of nodes and edges just
like the second variant of the ER model, but each edge is added one
at a time with a certain probability akin to the first variant. At
each step of graph generation, we select a pair of unconnected nodes
uniformly from all such node pairs, and we connect them with an edge
with a given probability. This process is repeated until the required
number of edges $m$ have been added. If the probability for adding each
edge was always fixed, then this model would be the same as the second
variant of the ER model. Instead, in our model, there is a base probability
$p_{0}$ for edge placement, but two additional mechanisms are included
to allow this probability to vary. The first mechanism is triadic
closure: should connecting the two selected nodes with an edge
close a triad, then the probability will be increased by $p_{1}$
over the base probability for adding the edge. We dub the second mechanism
``triadic closure plus'': should connecting the two selected
nodes with an edge close more than one triad, then the probability
will be further increased by $p_{2}$ for each potentially closed
triad in excess of one. Should the sum $p_{0}+p_{1}+t_{c}p_{2}$,
where $t_{c}$ is the number of closeable triads, exceed 1, it
will be interpreted as 1. Figure \ref{fig_3} illustrates these two additional
mechanisms. Though this model is fairly simple, each mechanism
can be motivated by consideration of the domain of application. In a friendship network, the
first mechanism corresponds to the idea that two people are more likely
to become friends if they have a mutual friend, while the second mechanism
further increases the likelihood for each additional mutual friend.
As such, these mechanisms can also be related to the so-called weak
ties hypothesis \citep{granovetter1973strength}, and it has been
shown that higher proportions of shared friends are associated with
greater tie strengths in large-scale communication networks \citep{onnela2007structure}.

\subsection{Model Selection}

We next assess the performance of our model selection framework through
simulation. We use various candidate algorithms, both on their
own and as members of the SL candidate algorithm library, to select between the full model
with both triadic mechanisms vs. the submodel with only the first
triadic closure mechanism. Networks generated from both models have
100 nodes and a base edge probability of $p_{0}=0.3$. The probability
of edge placement increases by $p_{1}=0.1$ for the first closed triad
and by $p_{2}$ for every subsequent triad, and we vary $p_{2}$
over the values 0.005, 0.01, 0.03, and 0.05. We vary the number of
edges for the two models over the values
500, 1000, and 2000. For a given number of
edges, the variation of $p_{2}$ from 0.005 to 0.05 makes the differences
caused by the additional mechanism easier to detect. For a given value
of $p_{2}$, the variation of the number of edges from 500 to 2000
means that there will be more opportunities for the additional mechanism
to manifest itself, also making it easier to detect. The simulation
studies iterate through each combination of values of $p_{2}$ and
edge count to see the interplay between the two factors.

In the simulations, SL used a library of three candidate algorithms:
$k$-nearest neighbors (KNN), support vector machine (SVM), and random
forest (RF). These three candidate algorithms are chosen largely for their ability
to handle collinear predictors, which are often present in summary
statistics for networks. Given an observed sample for classification,
KNN determines which $k$ samples in the training data
are closest to the observed sample, based on some distance measure
in the predictor space, a common choice being the Euclidean distance.
The predicted class is the most frequent class among the $k$-nearest
neighbors. Unlike KNN, which essentially formulates a new decision
rule for each observation, SVM seeks to formulate a single decision
rule by separating the space of the predictors
with a set of hyperplanes that segregate the space class-wise.
Once the class-wise segregation of the predictor space is complete,
a new data point is classified based on the class label of the subset
of the predictor space it falls in. Lastly, RF seeks to create a set
of decision trees (the ``forest'') from the training data in order
to arrive at the final prediction. To build each tree, a bootstrap
sample of the training data is taken to form the root. Then, at each
node of each decision tree, a subset of the predictors is selected at random, and a ``best'' split
is determined for these predictors in order to form its daughter nodes.
Typically, the quality of the split is measured by the amount of homogeneity
in each daughter node. Given an observation, each tree in the forest is traversed
and gives a class label for the observation based on the most frequent class among
the samples in the leaf of the tree where the observation is assigned.
An observation is then classified
by the forest as the mode of all the tree-wise decisions.

In addition to the library of candidate algorithms, choosing appropriate
predictors is an important task. As discussed in the
previous sections, sufficient summary statistics are difficult to
obtain in all but the trivial mechanistic network models, and one
should aim to use summary statistics that characterize
the differences between the candidate models. In the simulations presented here,
there are five summary statistics chosen as predictors. The first
predictor is the triangle (closed triad) count, which is an obvious choice since
the additional mechanism in the full model will favor edges that close
multiple triads. The second is the average local clustering coefficient
over all nodes. The local clustering coefficient of a node is a measure
of how close its neighbors are to forming a complete subgraph by themselves,
i.e., having every possible edge between any two neighbors. If the
addition of an edge between nodes $a$ and $b$ will close multiple
triads, then, without loss of generality, from the point of view
of node $a$, the addition of the edge would mean the addition of
a single neighbor, $b$, and the addition of multiple edges amongst
$a$'s updated set of neighbors between $b$ and the shared neighbors
of $a$ and $b$. In scenarios with lower total edge counts, where
the degree of either $a$ or $b$ is likely to be low, this could
lead to a potentially large change in the local clustering coefficient.
Lastly, the additional mechanism is a rich-get-richer scheme in
terms of the degree of a node, since the more closable triads
a pair of nodes has, the higher their degrees, which 
leads to a higher probability of both getting an increase to their
degrees with an additional edge. Thus, this mechanism is likely to
affect the degree distribution of the network. As a proxy to the full
degree distribution, the three quartiles (25\%, 50\%, 75\%) of the
degree distribution are included as predictors.

The mechanistic model was coded in Python using NetworkX. The training of SL, with
a 5-fold cross-validation, was done with the R package SuperLearner,
which contains wrappers for the chosen candidate algorithms. The parameters
for the candidate algorithms themselves are kept at the default values
($k=10$ for KNN; $N_{tree}=1000$, min terminal node size = 1 for RF; $\nu$-classification with $\nu=0.5$, radial kernel for SVM; see SuperLearner package documentation for more details).
For a given combination of edge count and value for $p_{2}$, we generated
10,000 training samples from the full model (with both triadic closure mechanisms)
and the submodel (with only the first triadic closure mechanism) as training data.
Rather than using a separate sample to assess performance,
the $AUC$ of SL as well as that of each candidate algorithm were estimated
via a 10-fold cross-validation. Note that this is a separate stage
of cross-validation from that of the training of SL itself. First,
the training data is partitioned into 10 validation sets, which are
used for performance cross-validation. Then, for each of the 10 performance
validation sets, a 5-fold cross-validated SL is trained on the union
of the remaining 9 performance validation sets, and then used to predict
the model index for the given performance validation set. The $AUC$
measure is computed for each of the 10 performance validation sets
and averaged.

\begin{figure}[!h]
\centering\includegraphics[width=\textwidth]{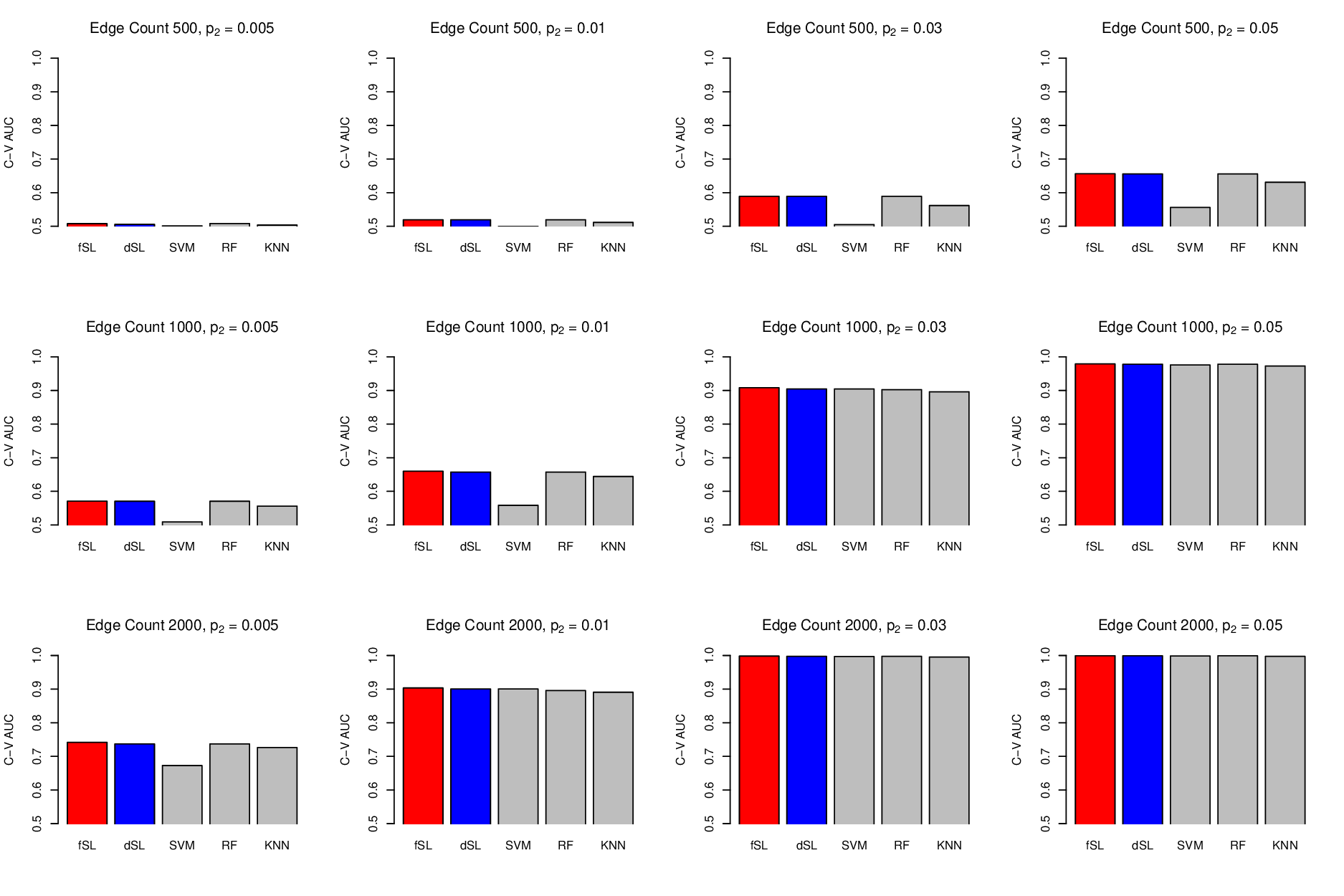}
\caption{Cross-validated $AUC$ for each method:
full SL (fSL, red), discrete SL (dSL, blue), support vector machine
(SVM, grey), random forest (RF, grey), $k$-nearest neighbors (KNN,
grey), in each simulation scenario.}
\label{fig_4}
\end{figure}

The cross-validated estimate of the $AUC$ for the full SL, the discrete
SL, and each candidate algorithm for each simulation scenario is summarized in Fig. \ref{fig_4} with numerical results in Table \ref{table_1}.
In general, performance deteriorates as the value of
$p_{2}$ decreases, which is no surprise as the effect of the additional
mechanism diminishes as $p_{2}$ gets smaller, and thus the two models
become harder to distinguish. In contrast, performance improves as the edge count
increases, as the mechanism has more opportunities to manifest itself
with more edges. In general, the model selection problem is easier
for larger networks.

\begin{table}[!h]
\caption{\textit{Cross-validated AUC from the simulations}}
\label{table_1}
\centering
\begin{tabular}{lrrrr}
$p_{2}$ & $0.005$ & $0.01$ & $0.03$ & $0.05$\tabularnewline
\hline 
\multicolumn{5}{c}{$EC=500$}\tabularnewline
\hline 
fSL & 0.50786 & 0.51941 & 0.58896 & 0.65630\tabularnewline
\hline 
dSL & 0.50575 & 0.51950 & 0.58911 & 0.65585\tabularnewline
\hline 
SVM & 0.50097 & 0.49936 & 0.50504 & 0.55592\tabularnewline
\hline 
RF & 0.50826 & 0.51950 & 0.58911 & 0.65585\tabularnewline
\hline 
KNN & 0.50353 & 0.51158 & 0.56172 & 0.63093\tabularnewline
\hline 
\multicolumn{5}{c}{$EC=1000$}\tabularnewline
\hline 
fSL & 0.57071 & 0.65993 & 0.90814 & 0.97895\tabularnewline
\hline 
dSL & 0.57055 & 0.65738 & 0.90445 & 0.97798\tabularnewline
\hline 
SVM & 0.50897 & 0.55834 & 0.90445 & 0.97586\tabularnewline
\hline 
RF & 0.57055 & 0.65738 & 0.90248 & 0.97798\tabularnewline
\hline 
KNN & 0.55596 & 0.64410 & 0.89605 & 0.97275\tabularnewline
\hline 
\multicolumn{5}{c}{$EC=2000$}\tabularnewline
\hline 
fSL & 0.74179 & 0.90348 & 0.99839 & 0.99947\tabularnewline
\hline 
dSL & 0.73703 & 0.90067 & 0.99769 & 0.99917\tabularnewline
\hline 
SVM & 0.67271 & 0.90067 & 0.99709 & 0.99879\tabularnewline
\hline 
RF & 0.73703 & 0.89573 & 0.99769 & 0.99917\tabularnewline
\hline 
KNN & 0.72623 & 0.89077 & 0.99540 & 0.99751\tabularnewline
\hline 
\end{tabular}
\justify{Cross-validated AUC from the simulations for
the full SL (fSL), discrete SL (dSL), support vector machine (SVM),
random forest (RF), $k$-nearest neighbors (KNN) over different edge counts (EC) and values
of $p_{2}$, the increase in probability for adding an edge for each potentially closed triad 
in excess of one. Discussion for the ordering of performance for EC = 500 appears 
later in this section.}
\end{table}

The simulation results seem to support the oracle properties of 
the discrete and full SL. Indeed, in most scenarios, the discrete
SL has the same cross-validated $AUC$ as the best performing candidate
algorithm, with the full SL performing a little better, as expected.
In these scenarios, the ordering of the candidate algorithms by performance
is likely the same across each fold in the cross-validation. Thus,
the discrete SL always picks the same candidate algorithm in each
fold and has the same performance as the best candidate algorithm
averaged across all folds. The full SL, in this case, takes a convex
combination of the candidate algorithms in each fold and performs
at least as well as, and likely better than, the best performing candidate
algorithm in each fold. When averaged across the cross-validation
folds, the full SL clearly performs better, as evidenced by the simulations.
Since it selects the candidate algorithm with the best cross-validated
performance by design, the discrete SL's oracle property is mediated
through the optimal candidate algorithm performing the best asymptotically.

Still, there are a few scenarios where either or both of the discrete
and full SL perform slightly worse than the best performing candidate
algorithm. This occurs with edge count 500 and $p_{2}\in\left\{ 0.005,0.01,0.03\right\} $.
In these scenarios, there are several candidate algorithms that are
quite close in performance, and the ordering of their performance
is likely not constant across the cross-validation folds. In this case, the discrete
SL picks different candidate algorithms across the folds and the $AUC$
averaged across the folds may be worse than that of the best candidate
algorithm. The full SL on the other hand weights different candidate
algorithms differently across the folds, and the cross-validated
$AUC$ can also end up worse than the best candidate algorithm.

In the scenarios where either or both of the discrete and full SL
perform worse than the best candidate algorithm, they both still perform
very close to the best candidate algorithm, with difference in $AUC$
only of the order of $10^{-3}$ or less. Since the oracle properties
for both SLs are asymptotic results, some small deviations from the asymptotic
behavior would be expected for finite samples. In the limit, the
ordering of the performance of the candidate algorithms are likely
to be constant across all folds, and both SLs are expected to perform
no worse than the optimal candidate.

The best performing candidate algorithm varies between SVM and RF
across the scenarios, which is not something known \textit{a priori},
but both versions of SL are able to closely match or beat the best
candidate algorithm. Although the differences between the best and
worst performing candidate algorithms in the scenarios visited are
not too great (biggest difference is about 0.1), it is still easy
to see the manifestation of the oracle property, which would be even
more valuable in scenarios with greater differences. Thus, when one
particular candidate algorithm outperforms the rest, one can be confident
that the discrete SL will select it, and the full version will weight
it heavily. In cases when one is unlikely to fully grasp the significance
of the various summary statistics as predictors or what learning algorithm
would work best due to the complex nature of the data, it would be
ideal to consider multiple candidate algorithms paired with different
combinations of predictors and find an optimal mix. This makes SL
well suited for model selection for mechanistic network models.

One interesting note is that the magnitude of the difference in performance
between the full SL and the best performing candidate algorithm (random forest
most of the time) is not too great. This reflects results from other papers in the
literature that use the SL specifically for classification \citep{pirracchio2014improving,pirracchio2015mortality,petersen2015super}.
Despite the relatively modest gain, we believe the flexibility the SL
offers is important for this setting. In addition to the choice of predictors and
candidate algorithms, the choice of the loss function allows one
to define the performance metric, which may not be possible with the candidate
algorithms. The oracle property also takes care of the analog of
the multiple comparison problem in the model selection setting by allowing the
user to gain information from multiple candidate algorithms yet make only one ``selection''.

We implemented the proposed procedure for quantifying the uncertainty
in the selected model on the simulated data. We assessed two scenarios,
an easy one, with edge count 2000 and $p_{2}=0.005$, and a difficult
one, with edge count 500 and $p_{2}=0.005$. For each scenario, we
first defined the $\widehat{M}^{*}$ as 1, or the ``positive'' model,
if the score given by the cross-validated SL was $>0.5$, and 0 or the ``negative''
model otherwise. We defined $W$ as above and built 
the regression model for $P(\widehat{M}^{*}=Y|\boldsymbol{X})$
with a new SL with the appropriate loss function in order to maximize
the corresponding binomial likelihood
\[\prod_{i}P(\widehat{M}_{i}^{*}=Y_{i}|\boldsymbol{X}_{i})^{W_{i}}(1-P(\widehat{M}_{i}^{*}=Y_{i}|\boldsymbol{X}_{i}))^{1-W_{i}}.\]
The new SL used the same library of candidate algorithms as before.
The average value of $P(\widehat{M}^{*}=Y|\boldsymbol{X})$
for the two scenarios were 0.71 for the easy scenario and 0.53 
for the difficult one. The average values for the easy scenario is
noticeably higher than 0.5, while that of the difficult scenario is
very close to 0.5 and thus hardly better than a random guess. These results 
are compatible with our AUC simulation results.

\section{Protein Interaction Network}

We apply our model selection approach to the yeast (\textit{S.cerevisiae}) protein-protein
interaction (PPI) network obtained from the database of interacting proteins (DIP)
\citep{salwinski2004database}. We use models from
two publications \citep{hormozdiari2007not,schweiger2011generative} that
fit different duplication divergence models \citep{sole2002model,pastor2003evolving} to the same data set.

Duplication divergence models are used in systems biology
for modeling PPI networks. The model fits in
\citet{hormozdiari2007not} and \citet{schweiger2011generative} both contain the
same mechanisms, but different seed networks and parameter values. The models 
generate network realizations by growing a seed network until the requisite number of 
nodes $n$ is reached. At the beginning of each step in network generation,
an existing node is chosen uniformly at random and duplicated by adding a new
node and connecting the new node with each neighbor of the chosen existing node. Each edge
connected to the new node is then removed with probability $1-p$. Finally, an edge
is added between the new node and any existing node at the start of the $t$th time step
with probability $r/n_t$, where $n_t$ is the number of nodes at the beginning of the $t$th step.

The two model fits use different parameter values and different seed networks.
The fit from \citet{hormozdiari2007not} uses a seed network of 50 nodes.
This seed network is constructed by connecting two cliques (complete graphs) of 7 and
10 nodes with several edges. Then another 33 nodes are connected with randomly chosen nodes from
the two cliques. The fit from \citet{schweiger2011generative} uses a seed network of 40 nodes.
The seed network is generated
from an inverse geometric model, where coordinates in $\mathbb{R}^{2}$ are generated
for each node. Then, each pair of nodes with Euclidean distance between corresponding
coordinates greater than threshold $R=1.5$ are connected. Each dimension of the coordinates
is generated independently from the standard normal distribution.

To apply our model selection approach, we generated network realizations from each model to fit 
the SL. Note that the results from \citet{hormozdiari2007not} and
\citet{schweiger2011generative} are based on different older releases of the DIP data, while our analysis
only uses the latest data release. Thus, we will only use the seed networks
(these are independent of the empirical network) from the respective
models fits, but not their corresponding parameter estimates. Furthermore, rather than refitting
both models, we take the approach from \citet{RF_ABCpudlo2015reliable}
and draw parameter values from a prior (uniform over the unit square), then generate a network
realization from each model for each parameter value as the training data.

The SL here uses the same candidate algorithms we used in the previous section. The predictors
 here are the criteria of fit from both papers. Betweenness centrality of a node $v$ is the number
of shortest paths across all pairs of nodes that pass through $v$, while closeness centrality of $v$ is the inverse
of the sum of the lengths of all shortest paths from $v$ to all other nodes. The $k$-hop reachability of a node
$v$ is the number of distinct nodes that can be reached in $k$ edges or less. \citet{hormozdiari2007not} use 
these measures to compare the overall connectivity of networks generated from their model with that of the
empirical PPI network. We use the means of the two centrality measures as well as that of $k$-hop reachability up to
$k=6$ as predictors.

Bicliques are subgraphs between two distinct sets of nodes where
every edge between the two sets exists. A biclique is maximal if it is not a strict subgraph of another biclique.
\citet{bu2003topological} notes that large bicliques exist in the yeast PPI network, and they are used to infer
protein function as well as to find binding motifs \citep{li2006discovering}. \citet{schweiger2011generative}
based the fit of different models on maximal bicliques with distinct sets of nodes with at least 2 nodes.
For consistency, maximal bicliques are represented
as ordered pairs of the number of nodes in the two distinct sets, with the first element being less than or equal
to the second. This can be cast as a bivariate distribution. To use as predictors, we summarize the maximal
biclique distribution of a network with the total number of maximal bicliques, a norm of the distribution of
maximal bicliques, the mean and variance of each dimension of the bivariate distribution, as well as the Pearson
and Spearman correlation between the two dimensions. The norm is computed as
$\sqrt{\sum_{2\le n \le m}\left( \log_{10}\left( p_{(n,m)}+1 \right) \right)^2}$, where $p_{(n,m)}$ is the proportion
of maximal bicliques with sets of nodes with size $n$ and $m$. This choice is motivated by the distance used in
\citet{schweiger2011generative}.

For the purposes of model selection and without loss of generality, we assigned the model of \citet{hormozdiari2007not} as the 0 or
``negative model'' and that of \citet{schweiger2011generative} as the 1 or ``positive'' model.
We trained 3 SLs: the first is based only on the predictors from \citet{hormozdiari2007not},
the second is based only on those from \citet{schweiger2011generative}, while the last one is based on those from
both papers. Note that we use the definitions of the predictors from the two papers as given and compute their values 
from the latest release of the data.
Predictably, the first two trained SLs evaluated at the corresponding predictors for the PPI network
returned opposing results, each favoring the model its corresponding predictors were from. The first SL
returned a score of approximately 0.00629 and very strongly favors model 0, while the second returned a score of approximately
0.788 and  strongly favors model 1. Both scores are closer to their respective extremes than 0.5, suggesting at least moderate
evidence for their respective models. The third SL returned a score of approximately 0.658, which favors model 1, but is
closer to 0.5 than either of the scores from the previous 2 SLs. The results from using both sets of summaries thus favor the model of \citet{schweiger2011generative} for the most current (\textit{S.cerevisiae}) PPI dataset.

These results exemplify the difficulty with model selection for mechanistic models where predictors are insufficient 
summaries of the data. Each of the summaries used as predictors are reasonable criteria for fit as presented in their respective
papers, but they can lead to conflicting results depending on the subset of predictors used, as with the first two trained SLs. It can therefore  
be difficult to compare these results.
In general, with classification methods such as random forest or SL based on such learning algorithms,
using a larger set of predictors typically does not reduce performance, so it is reasonable to use a larger set of predictors. However,
one can always come up with new summaries. Thus, a natural question is when to stop expanding the set of predictors/summaries.

One means of assessing the usefulness of a predictor is through an importance metric. As a reference, we report in Table \ref{table_2} the random forest
feature importance \citep{breiman2001random} for each predictor in the three selections specified in the previous paragraphs.
In the first selection (SL1), based only on predictors from \citet{hormozdiari2007not}, the two mean centrality measures as well as 6-hop reachability had
the three highest feature importances. In the second (SL2), based only on predictors from \citet{schweiger2011generative}, the norm of the maximal
biclique distribution, as well as the mean and variance of the second dimension of the distribution, had the highest feature importances. When the two sets
of predictors are combined in SL3, the top three predictors from \citet{schweiger2011generative} retain high importance, while
those from \citet{hormozdiari2007not} all have lower importance with only betweenness centrality retaining moderate importance. In this
example, it seems the predictors from \citet{schweiger2011generative} are more discerning. The example also shows that the feature importance
can be a useful tool in distilling a large set of available predictors to build a classifier for model selection.

\begin{table}[!h]
\caption{\textit{Feature importance in the three selections based on different sets of predictors}}
\label{table_2}
\centering
\begin{tabular}{lrrlrr}
Predictor \citep{hormozdiari2007not} & SL1 & SL3 & Predictor \citep{schweiger2011generative} & SL2 & SL3\tabularnewline
\hline 
\textbf{Bet. cent.} & 0.431 & 0.113 & Total bic. & 0.012 & 0.007\tabularnewline
\hline 
\textbf{Clo. cent.} & 0.201 & 0.045 & \textbf{Norm bic.} & 0.229 & 0.191\tabularnewline
\hline 
1-hop & 0.059 & 0.004 & Mean 1-d & 0.060 & 0.028\tabularnewline
\hline 
2-hop & 0.059 & 0.005 & \textbf{Mean 2-d} & 0.264 & 0.230\tabularnewline
\hline 
3-hop & 0.033 & 0.003 & Var 1-d & 0.040 & 0.016\tabularnewline
\hline
4-hop & 0.029 & 0.002 & \textbf{Var 2-d} & 0.339 & 0.311\tabularnewline
\hline
5-hop & 0.054 & 0.005 & Pearson & 0.040 & 0.014\tabularnewline
\hline
\textbf{6-hop} & 0.134 & 0.021 & Spearman & 0.017 & 0.005\tabularnewline
\hline
\end{tabular}
\justify{Feature importance in the three selections
that are based on different sets of predictors: selection 1 (SL1) is based only
on those from \citet{hormozdiari2007not}, selection 2 (SL2) is based only on
those from \citet{schweiger2011generative}, and selection 3 (SL3) is based
on both sets. The three predictors with the highest importance in SL1 and SL2 are bolded.}
\end{table}

\section{Discussion}

Network models are widely used in many domains, and mechanistic models
allow one to easily incorporate domain knowledge. However, due to
the intractability of the likelihood of the typical mechanistic network
models, likelihood-based model selection methods are not feasible.
We propose a procedure that combines the Super Learner (SL) framework
with the data generation of Approximate Bayesian Computation (ABC),
allowing one to leverage the ease of generating data from mechanistic
models via forward simulation, while quantifying uncertainty in the
selected model.

While ABC provides one viable means for mechanistic network model
selection, an accurate ABC posterior requires the knowledge of sufficient
statistics \citep{barber2015rate,lintusaari2017fundamentals},
which are typically difficult to find in the case of intractable
likelihoods. The use of insufficient statistics and non-zero threshold for the
distance, as is commonly done in ABC, can make the approximation of the posterior
distribution of the model index worse. As an alternative to
ABC for likelihood-free model selection, we proposed to use SL for model
selection while borrowing the generation of pseudo-data from ABC.
While it still suffers from the lack of sufficiency, our approach aims
to make the best use of available predictors.

With training data readily generated from each candidate model, SL
seeks to build an optimal prediction algorithm from a library of candidate algorithms.
In this case, it seeks to build an optimal classifier from candidate
algorithms to best discriminate between the available models with
the given predictors that are not necessarily sufficient. One is
unlikely to know what pairing of classifiers and predictors will
perform well, but with SL one does not need to make this choice as
SL will try to build the optimal classifier with all that is given.
However, this does not mean that the quality of the predictors does
not matter. The better the predictors are at characterizing the differences
between the candidate models, the better SL performs. Though the ability
to characterize the differences likely correlates with sufficiency,
we cannot use sufficiency as a criterion for choosing
the predictors. For mechanistic models, one can, and should, apply
domain knowledge to select predictors.

Our procedure assumes that one has well-defined candidate models. Bayesian models require a corresponding prior distribution, whereas 
non-Bayesian models require the calibration of model parameters. 
For such calibration, it is important to consider
which characteristics of the networks are unlikely to be affected by the different mechanisms
of the candidate models, and then to calibrate the parameters of each
candidate model based on these characteristics of the observed network.
This will hopefully allow the differences in the candidate models
to more clearly manifest themselves and to ensure that the data generated
from each candidate model are similar to the observed data. Ideally, this also means that the data generated from the
true model will match the observed data closely. However, due to the nature
of the likelihood, calibration of model parameters in the latter scenario is not straightforward.
This is another situation where likelihood-free methods such as ABC are appropriate.

Finally, it is possible to approach the model selection problem in a very different way: one can
turn the model selection problem into a density estimation problem in the summary statistic space.
Briefly, one can use kernel density estimation, or similar methods, to estimate the probability density
function of the network summary statistics conditional on the model used to generate the networks.
By introducing a prior probability on the model index, one can then use Bayes' theorem to arrive at
the model posterior probability conditional on the observed summaries. The choice of summary statistics
can then be guided by the idea that informative summaries, associated with different mechanistic models,
should result in minimal common support in the estimated model-specific densities in the summary statistic
space. This approach would not alleviate the problem of insufficient statistics, but rather than having to
specify a threshold for the distance metric, one would now have to specify the precision (width) of the
kernels used in density estimation. Density estimation is known to be difficult in high dimensional spaces,
but it might be possible to use a heuristic for proposing small sets of summaries such that the
dimensionality of the summary statistic space does not become prohibitive. We leave further
development of this idea for future work.

\section*{Funding}
This work was supported by the National Institutes of health [1DP2MH103909-01 to S.C. and J.P.O., 5U01HG009088-02 to S.C., U54GM088558-09 to S.C., 5R37AI051164-12 to J.P.O., 1R01AI112339-01 to J.P.O., U54GM088558-06 to J.P.O.]; and the Swiss National Science Foundation [CR12I1\_156229 to S.C. and A.M., 105218\_163196 to A.M.].

\section*{Acknowledgements}
We would like to thank Dr. Laura Balzer for her generous help on the Super Learner.

\bibliographystyle{abbrvnat}
\bibliography{paper_3_ref_rev}

\end{document}